%% file: paper.tex
\documentclass[conference,letterpaper]{IEEEtran}
\IEEEoverridecommandlockouts
\usepackage[nosort]{cite}
\usepackage{url}
\usepackage{amsmath,amssymb,amsfonts}
\usepackage{algorithm}
\usepackage{algpseudocode}
\usepackage{needspace}
\usepackage{graphicx}
\usepackage{textcomp}
\usepackage[utf8]{inputenc}
\usepackage{array,tabularx,booktabs,multirow}
\usepackage[table]{xcolor}
\definecolor{NvidiaGreen}{HTML}{76B900}
\colorlet{TableHeader}{NvidiaGreen!15!white}
\colorlet{TablePositive}{NvidiaGreen!25!white}
\usepackage[hidelinks]{hyperref}
\newcolumntype{Y}{>{\raggedright\arraybackslash}X}
\newcolumntype{P}[1]{>{\raggedright\arraybackslash}p{#1}}
\algrenewcommand\algorithmicrequire{\textbf{Require:}}
\algrenewcommand\algorithmicensure{\textbf{Ensure:}}

\begin{document}
\bstctlcite{IEEE:BSTcontrol}
\title{One Megakernel is Not Enough: Predicting Where to Cut for Streaming Inference}
\author{
\IEEEauthorblockN{Aditya Chatterjee}
\IEEEauthorblockA{\textit{NVIDIA}\\Bangalore, India\\adichatterje@nvidia.com}
\and
\IEEEauthorblockN{Anand Joseph X M}
\IEEEauthorblockA{\textit{NVIDIA}\\Hyderabad, India\\anajoseph@nvidia.com}
\and
\IEEEauthorblockN{Sanjay Singh Chauhan}
\IEEEauthorblockA{\textit{NVIDIA}\\Pune, India\\schauhan@nvidia.com}
}
\maketitle
\begin{abstract}
A megakernel is a large monolithic kernel that fuses multiple kernels/operators into a single persistent kernel. We show that the best configuration is a composition of multiple megakernels with no overlapping conventional graph component, or a fully conventional graph. We present a framework that selects megakernel regions in a streaming inference model from conventional-graph profiles and standalone kernel benchmarks, without running any candidate design end to end. Across NH-TTS, Qwen3-TTS and Llama-1B, the best composition ranges from zero to four megakernels at different batch sizes. We introduce four model categories (megakernel-stable/ receding/ advancing/ averse) that capture how a model's best megakernel composition changes with batch size. Using NH-TTS model on H100 NVL GPU, we demonstrate that preserving three specialized megakernels can improve both latency and throughput despite introducing additional kernel boundaries.
\end{abstract}

\section{Introduction}
Time to First Audio (TTFA)/Time to First Token (TTFT), Inter-Token Latency (ITL)/Inter-Chunk Latency (ICL) and Throughput (RTFx)/Realtime Factor (RTF) are the key metrics for real-time serving of streaming inference models. So long as the RTFx is greater than 1, the overall latency of the response is dictated by the TTFA / TTFT. Minimizing TTFA/TTFT is critical in applications that support bidirectional streaming (input and output). Being able to serve requests at the maximum batch size while keeping these metrics within acceptable limits is critical for real-time serving.

A megakernel fuses multiple operators into one persistent kernel whose thread blocks stay resident on the GPU and hand-over work to one another without returning to the host. Large Language Models (LLMs) operate in two phases per request - prefill (run once) and decode (run autoregressively token by token). Existing megakernel designs usually fuse the decode phase and leave prefill to conventional execution or out of scope. In~\cite{cheng2026mpk} up to 1.7\texttimes{} lower per-token latency than vLLM~\cite{kwon2023vllm} and SGLang~\cite{zheng2024sglang} is reported for small-batch LLM decoding. Multi-modal models like text-to-speech (TTS) operate in multiple phases. A TTS model runs prefill once per request, an autoregressive producer once per step, a local transformer N times per step and an audio codec once per chunk of frames. These phases run at different rates and prefer different thread-block shapes. These phase differences make fusion boundaries a performance choice. Each cut decides which phases share one launch configuration, how long thread blocks stay held and how the phases share the GPU.

\vspace*{-8.94pt}
We evaluate an upcoming autoregressive TTS model with a Nemotron-H backbone~\cite{nvidia2025nemotronh} denoted as NH-TTS in this paper. Figure~\ref{fig:nh_tts_layout} shows the six representative NH-TTS layouts at batch size 32. For NH-TTS at batch 32, Figure~\ref{fig:mk_perf} shows that the three-megakernel design publishes the first audio 4.05\texttimes{} sooner than the conventional TensorRT~\cite{nvidia2026tensorrt} pipeline and 21.4\% sooner than a single megakernel for the whole model. Across 13 combinations of model and batch size on three frameworks, the best tested design uses anywhere from zero to four megakernels. Qwen3-TTS~\cite{hu2026qwen3tts} needs four up to batch 32 and none from batch 48, where the same cuts make it slower. Guessing is costly: for NH-TTS and Qwen3-TTS, the next-best layout is 3 to 29\% slower than the best.

\vspace*{8.94pt}
\input{figures/original/figure-1-float}

Exhaustively implementing and timing compositions is costly: our NH-TTS search contains 81 candidates. We address first-publication-aware selection among supported implementations by combining compatible service measurements with persistent resource lifetimes and streaming publication dependencies.

We present a boundary planner that ranks supported compositions using conventional-graph profiles, operator benchmarks and source-declared implementation contracts. These contracts specify geometry, storage, compiled-entry membership and synchronization. The planner combines their costs with streaming dependencies to select boundaries and sharing policies without timing candidate compositions end to end.

We evaluate the planner on H100 NVL GPU with three models on three frameworks: NH-TTS (TensorRT), Qwen3-TTS (vLLM~\cite{kwon2023vllm}) and Llama (PyTorch~\cite{ansel2024pytorch2}). We show that the planner selects the measured best design in all 13 model-batch cases.

\section{Background}
In codec-based autoregressive TTS, prefill initializes the request context. The acoustic-token generator (producer) emits frames of discrete acoustic codes and the neural codec decoder (consumer) converts buffered frames into PCM waveform samples. Producer and consumer denote scheduling roles relative to the acoustic-code buffer. A frame denotes one codec time step; a chunk groups frames for decoding and publication.

\subsection{Streaming Inference and First Publication}
A model's streaming schedule states when output may be published. A TTS model may wait a few steps before speech starts, decode its first chunk once enough frames exist and publish that chunk only after one more producer step shows that the stream continues. First publication occurs when the first audio chunk or token becomes available. The planner minimizes a deterministic first-publication score.

\subsection{Conventional Graph Execution}
Inference frameworks such as TensorRT, vLLM and PyTorch run a model as a sequence of kernels, typically one per operator or small fused group. Each kernel launches a grid of thread blocks called cooperative thread arrays (CTAs) which the hardware spreads over the GPU's streaming multiprocessors (SMs). H100 NVL GPU has 132 SMs. A kernel boundary is a barrier: the next kernel starts only after the previous one completes. CUDA graphs record a sequence of launches and replay it with one host call which removes host launch overhead. At small batch size, the kernels are short so host launches, barriers and host synchronization take a large share.

\subsection{Megakernels}
A megakernel is a persistent kernel~\cite{cheng2026mpk}. It is launched once with no more thread blocks than the GPU can hold at the same time and those blocks stay resident and run many operators in turn from an instruction stream. At a scheduler level, this adds multiple advantages like memory prefetch and tile execution across original kernel boundaries leading to better GPU utilization~\cite{spector2025nobubbles}.

Fusion also adds costs. All operators inside a megakernel share one launch configuration lacking specialization and higher register pressure leading to register spilling~\cite{spector2025wholegpu}.

\subsection{Megakernel Designs and the Boundary Selection Problem}
When a model is split into several regions, the regions can share the GPU serially, each taking the whole GPU in turn; side by side on a dynamic or static split of thread blocks. A design makes three choices: a partition of the model's phases into contiguous regions, a backend for each region (conventional or megakernel) and a sharing policy.

The boundary-selection problem is to choose a supported composition using conventional-execution profiles and operator benchmarks. The planner assigns each candidate a deterministic first-publication score and selects the candidate with the lowest score. This score is a ranking proxy, rather than a direct prediction of the latency distribution or its 95th percentile. We evaluate selection quality against the measured latency statistic defined for each model in Section~\ref{sec:evaluation}, within that model's implemented validation pool.

\section{Composing Megakernels: A Nemotron-H Based TTS Case Study}
Like a typical autoregressive TTS model, NH-TTS comprises prefill, acoustic-token generation and codec decoding. Each generation step executes the Nemotron-H backbone once and a local transformer N times, once per codebook, to produce one acoustic-code frame. The codec decoder converts chunks of frames into PCM samples. Under the evaluated streaming policy, first publication waits for both the decoded chunk and one further generation step.

The phases run at different rates, use different tensor shapes and thread-block sizes. Each boundary between them is a natural place to cut and each cut has consequences for the whole pipeline.

\subsection{Multiple Combinations}
We compare the conventional baseline of TensorRT with CUDA graphs against multiple native megakernel and conventional graph design combinations. Each design runs at batch size 32 on H100 NVL GPU. Three megakernels (Figure~\ref{fig:composition}) with overlap publish the first audio 4.05\texttimes{} sooner than conventional execution and 21.4\% sooner than one megakernel design at batch size 32. The ordering holds at other batch sizes (Bn denotes batch size n): at B64, 47.4 ms against 61.0 ms for one megakernel and 221.5 ms for conventional; at B1, 27.5 ms against 32.7 and 73.0 ms respectively.
\input{figures/original/figure-4-float}

\section{Automated First-Order Boundary Discovery from First Principles}
This section presents our method for identifying which kernel boundaries are worth preserving. It is first-order because it uses only measurements of the conventional graph and of standalone operators, never of a candidate design, and it works from first principles because it composes those measurements through explicit models of tiles, launches and shared GPU capacity.

The planner selects fusion boundaries and execution policies using conventional-graph profiles, standalone operator or step benchmarks, and source-declared implementation contracts. It first constructs a region graph (rGraph) and enumerates supported compositions. It then estimates implementation-specific service costs by constructing an intermediate tile graph (tGraph) and then, constructs an event graph (evGraph) for each composition. Simulating this graph yields a first-publication score used to rank candidates, without measuring candidate compositions end to end.

\subsection{Region graph (rGraph) and candidates}
We take a model with its runtime implementation as our input. We run the conventional graph implementation of the model using inference frameworks like TensorRT and capture the following information:
\begin{itemize}
\item Execution frequency: which model region runs once, every producer step or every audio chunk.
\item Workload characteristics: Arithmetic intensity, supported tile geometry, tensor sizes, memory traffic.
\item Dependencies: data and state updates which must finish before the current operation can proceed.
\item Publication: what computation must complete before the application can receive audio.
\end{itemize}

This results in Region Graph (rGraph) with three group types:
\begin{itemize}
\item Admission group: runs once per request
\item Producer group: run once per generated step
\item Consumer group: run once per output chunk
\end{itemize}

The region graph has one node per group with edges for data, state and publication dependencies. A region is a contiguous set of kernels executed by one backend:
\begin{itemize}
\item conventional execution with one kernel per operator
\item or a megakernel that runs the region in one persistent launch.
\end{itemize}

\noindent A cover partitions a fixed ordering of the groups into contiguous regions and assigns each region a conventional or megakernel backend. Before filtering by implementation support, $n$ groups yield $2\cdot3^{n-1}$ covers: the first group chooses either backend, and each subsequent group joins the preceding region or starts a new region with either backend. A candidate combines a cover with one supported execution policy.

An execution policy specifies how the regions of a cover share the GPU:
\begin{itemize}
\item Serial policy: regions run one at a time with the whole GPU.
\item Fixed split: producer and consumer regions are co-resident for the whole run with CTA budgets $X$ and $Y$.
\item Dynamic split: producer uses the whole GPU until the consumer's input is ready, then continues on $X$ CTAs beside the consumer's $Y$ CTAs.
\end{itemize}

The policy is declared from launch configurations that the implementations support and a policy is attached only to the cover where it applies. Overlap requires the producer groups to form one megakernel region and the consumer to be a megakernel. A dynamic split also requires them to be separate launches. A candidate is a cover together with one applicable policy.

A model's streaming schedule determines how much producer work precedes the first output. Three of its parameters matter:
\begin{itemize}
\item a delay of $d$ producer steps whose output is not yet emitted;
\item a first chunk of $c_1$ steps;
\item a lookahead of $l$ steps that shows whether the stream continues.
\end{itemize}

For NH-TTS, this exposes four groups:
\begin{itemize}
\item P: Prefill (admission)
\item B: Nemotron-H backbone (producer)
\item L: Local transformer (producer)
\item C: Codec (consumer)
\end{itemize}

For NH-TTS, the four ordered groups yield 54 serial covers. Fixed splitting applies to six covers: three with producer and consumer work inside one megakernel and three with separate producer and consumer launches. Dynamic splitting applies only to the latter three. With three supported CTA-budget pairs, these policies add $6\times3+3\times3=27$ candidates, giving 81 in total.

\subsection{Tile Graph (tGraph) and Service Table}
For each group, we build a tile graph (tGraph)~\cite{shi2023welder}:
\begin{itemize}
\item Each supported output tile becomes a task carrying its compute, weight and activation loads, output write, and staging storage.
\item Edges record which tiles supply each tile's inputs. The original kernel barriers are kept as a separate edge set.
\item Tile dependencies come from source-declared contracts, profiling kernels, grids and resource footprints. The planner checks these contracts against the observed profiles and calibrates tile service times to reproduce standalone execution. Operators without a tile decomposition remain opaque tasks occupying the modeled worker pool. Their reduced-budget costs use the CTA-wave approximation.
\end{itemize}

We schedule a tile graph with an event-driven list scheduler over a fixed pool of $W$ workers where $W$ is the CTA budget. A tile starts when its dependencies are complete and a worker with enough staging space is free. Memory transfers pass through a single FIFO memory queue and time advances to the next completion.

Two megakernel mechanisms act on this schedule:
\begin{itemize}
\item Removing the barrier edges lets ready tiles of the next kernel fill idle workers.
\item Weight prefetch lets a computing worker stage the weights of an upcoming tile.
\end{itemize}

The result is a service table: the estimated time of one execution for every group for each backend and CTA budget.

Service costs depend on the implementation contract of the candidate region. We write them as $S[g,\beta,W;\kappa]$, indexed by operation group $g$, backend $\beta$, CTA budget $W$, tensor shapes, and precision, under an implementation context $\kappa$. The context captures thread counts, block geometry, call and storage policies, compiled-entry membership, and synchronization requirements. Each record stores a measured or tile-simulated duration together with its provenance. Timing evidence is reused only under its declared transfer conditions: a record is reused only when its key matches exactly, and any candidate without compatible evidence is excluded. Consequently, the same group, backend, and CTA budget can incur different costs when the enclosing implementation changes.

Each candidate uses a source-supported implementation configuration. The cost model selects compatible operator measurements for that configuration; it does not infer arbitrary fused-entry compiler behavior from conventional timings alone. Thus, enclosing-entry context changes the service cost before composition scoring.

\subsection{Event Graph (evGraph)}
Each candidate is lowered to an event graph extending to first publication. Admission groups occur once, producer groups execute through step $h=d+c_1+l$, and the first consumer chunk becomes eligible after step $r=d+c_1$. Conventional regions retain their constituent operator launches. Each megakernel launch holds its CTA allocation until all tasks assigned to that launch finish, including periods when those tasks wait for dependencies. Dynamic splitting ends the full-GPU producer launch at step $r$ and continues with separate producer and consumer launches using budgets $X$ and $Y$. Dispatch nodes represent the associated launches.

Node durations come from the service table. Edges encode execution dependencies and chunk readiness. Publication waits for both the first consumer chunk and producer step $h$; consequently, $T_{\mathrm{pub}}=\max(F_{\mathrm{chunk1}},F_{\mathrm{producerh}})$, where $F$ denotes simulated completion time.

\subsection{Megakernel Impact Simulation}
The tile scheduler produces implementation-specific service costs. Algorithm~\ref{alg:simulate} uses these costs to simulate the event graph under launch-allocation and shared-service constraints. Table~\ref{tab:features} identifies the mechanisms modeled at each level.
\input{tables/original-features}

\subsection{Boundary Selection}
The planner exhaustively scores the supported candidate set and returns all minimum-score candidates. Candidates require compatible service-cost evidence; deadlocked schedules are infeasible. Ties are retained rather than resolved using validation measurements.

Algorithm~\ref{alg:select} describes the common scoring procedure; the model adapters use different representations of service costs. At NH-TTS batch 32, the planner selects $P/(B+L)/C$ : separate megakernels for prefill, the fused Nemotron-H backbone/local transformer and the codec with a dynamic producer/consumer split of 88/44 CTAs.

An illustrative calculation shows the tradeoff. Suppose chunk readiness occurs at 20 ms. Remaining producer and codec work take 4/6 ms serially and 6/8 ms under an 88/44 split. With serial / overlap transition costs of 0.1 / 0.2 ms and no bandwidth contention, publication scores are 20 + 4 + 6 + 0.1 = 30.1ms and 20 + max(6,8) + 0.2 = 28.2ms. Overlap therefore wins despite slower individual components.

\pagebreak[4]
\begingroup
\setlength{\intextsep}{0pt}
\enlargethispage{48pt}
\input{algorithms/original-simulate}
\endgroup
\newpage
\begingroup
\setlength{\intextsep}{0pt}
\input{algorithms/original-select-composition}
\endgroup
\vspace{12pt}

\section{What the Boundary Model Reveals about Composition}
The tile and event graphs predict and explain multiple observations on performance of Megakernels (and kernels in general) in a model execution environment.

\subsection{Observation 1: A faster kernel can make the pipeline slower}
In event graph, every occurrence is bound to a launch contract, which holds a CTA allocation from admission until its last task finishes. Co-running launches interact only through the joint simulation (Algorithm~\ref{alg:simulate}): they compete for CTAs and progress at the shared-service rate. The planner uses no measured co-run timings.

A slowdown with a faster kernel can happen due to changed contract that is more CTAs held or more memory traffic impacting the peer co-located kernels.

Based on experiments testing the observation, with 88 CTAs instead of 80, the producer megakernel in NH-TTS is $2.2\%$ faster alone yet beside the conventional codec, the pair takes $30.3\%$ longer because the codec loses eight CTAs; beside the codec megakernel, the pair gets $2.0\%$ faster.

\subsection{Observation 2: Colocation cost depends on the peer implementation}
A persistent launch retains its CTA allocation until completion. In Table~\ref{tab:latency_comparison}'s NH-TTS experiment, the conventional codec slows by 7.57\texttimes{} beside the producer megakernel, whereas the codec megakernel slows by 1.23\texttimes{}. The tested allocation therefore exhibits substantially different interference depending on the peer implementation.

\input{tables/original-codec}

\subsection{Observation 3: Serial cuts can be neutral under fixed contracts}
Excluding dispatch overhead and holding implementation and resource contracts fixed, inserting a serial boundary leaves the modeled device schedule unchanged. Under these assumptions, all eight all-megakernel partitions of NH-TTS predict 37.067 ms TTFA p95 on H100 NVL.

On H100 NVL GPU, the monolith and the two-megakernel design differ only by the prefill cut, with the same implementation and CTA split, and they measure within $0.5\%$ of each other at every batch size, as predicted. The monolith is slower than the serial three-megakernel design for a different reason: it runs every phase with 512-thread blocks, while the three-megakernel designs run Stage 0 with 256-thread blocks. The specialization, not the boundary, accounts for the gap.

\subsection{Observation 4: A megakernel's fixed cost decides when conventional execution wins}
Megakernel has fixed costs such as launches, host protocol and internal synchronization and cache residency across operations.

A megakernel's fixed cost does not shrink with the work it saves. Its lead comes from the overhead it removes and at large batches, conventional execution spreads that overhead over more work. Once the lead falls below the fixed costs, conventional execution publishes first.

On H100 NVL, Qwen3-TTS's four-region megakernel lead falls from 15.9 ms at B16 to 0.65 ms at B64 while its fixed costs stay at 0.7 to 1.1 ms so conventional graph wins at B48 and B64.

With NH-TTS, the lead with Megakernels keeps on increasing with batch size so conventional execution never wins at any tested batch size.

\section{Evaluation across models and frameworks}\label{sec:evaluation}
Using only execution of the conventional graph baseline and kernel benchmarks, our method picks the fastest megakernel composition.

\subsection{Setup}
We evaluate the models and batch sizes shown in Table~\ref{tab:evaluation_models}. Within each model, designs share weights, operator precision, input fixtures, generation limits, deterministic decoding and publication rules. The artifact records checkpoint hashes, software versions, text/context lengths and $(d,c_1,l)$. Before timing, each design is checked against an unfused reference on separate correctness fixtures: generated token or code indices and publication prerequisites must match exactly, and floating-point outputs must have relative L2 error at most 10-2. Failed designs are reported and excluded. Qualified designs are measured in balanced randomized order within matched GPU blocks.

\input{tables/original-models}

The pools contain the designs we could build and qualify, not every candidate: the planner ranks all 81 NH-TTS candidates, but only these are measured. Each design runs in its own process with exclusive use of one H100 NVL GPU. Each NVIDIA H100 NVL GPU has 94 GB of HBM3 memory with 3.9 TB/s peak bandwidth~\cite{nvidia2024h100nvl}; we use CUDA 13.2 and NVIDIA driver 595.84.

NH-TTS reports the mean of per-batch request p95 TTFA over 36 batches in the main campaign. Qwen3-TTS computes p95 over three cohort-level first-PCM publication times per run and averages 9 run-p95 values across 3 processes at B1/B48/B64, or twelve across four processes at B16/B32. Llama computes p95 over 30 repetitions per process and aggregates across three processes using the arithmetic mean. We discard two warmup cohorts per process for NH-TTS, Qwen3-TTS and Llama, before measurement. Ratio confidence intervals use 20,000 bootstrap resamples of four matched GPU blocks for NH-TTS and Qwen3-TTS B16/B32, or nine matched process/run units for Qwen3-TTS B1/B48/B64, with paired resampling across designs. Collecting the planner's inputs takes minutes: kernel benchmarks took 4.6 GPU-minutes for NH-TTS and 15.9 for Qwen3-TTS across four batch sizes.

We fixed the scoring rules and candidate-generation policy using a separate development set. Evaluation batch sizes and input fixtures were excluded from development. At each evaluation batch size, conventional and standalone component profiling supplied the permitted inputs. Candidate pools, service records, predicted scores and tie rules were frozen before candidate end-to-end timing. Validation results were not used to revise the model. Previously inspected cases were reported separately as development evidence.

\subsection{Selection accuracy on held-out cases}
The frozen planner uniquely selected the measured fastest design in 13/13 held-out model-batch cases. This establishes selection accuracy within the declared validation pools across the tested workloads on H100 NVL. It does not establish optimality over unmeasured compositions or transfer to other GPU architectures.

The planner ranks designs; it does not calibrate absolute time. Conventional designs, scored from a measured run, land within 5\% of validation. NH-TTS's one- and two-megakernel designs share one persistent implementation and land within 5\%; its three-megakernel designs share another and are over-predicted by 6 to 22\% (42.52 ms predicted for the selected B32 design, 35.99 ms measured). Llama-1B designs are over-predicted by up to 11\%, while Qwen3-TTS's four-megakernel design is under-predicted by 1.1 to 1.5 ms at B48 and B64. Despite these absolute errors, the predicted winner matches the measured winner in each validation pool. The closest decisions hold within their 95\% intervals: two megakernels are 8.1\% slower than four at Qwen3-TTS B1 (7.8 to 8.3\%), and four megakernels are 4.2\% (3.1 to 4.9\%) and 3.2\% (3.0 to 3.4\%) slower than conventional at B48 and B64.

Within NH-TTS's three-megakernel layout, the method also ranks the resource policies correctly: giving the producer and codec 80 and 52 thread blocks is $1.0\%$ slower than $88$ and $44$ at batch size $32$ and running the three megakernels one after another is $4.2\%$ slower than overlapping them at B64.

With identical service costs and candidate pools, the full planner matches $13/13$ measured winners; additive scoring and removal of overlap each match $11/13$, with maximum regret of $4.50\%$. Both misses occur at NH-TTS B32/B64, where the simpler selectors choose serial execution. The full planner instead accounts for chunk readiness, CTA allocations and concurrent completion. These results localize the benefit of event simulation to overlap-sensitive decisions; implementation-specific component costs explain selections where execution is serial.

For Llama-1B, the two-megakernel design runs prefill and decode as separate megakernels. It beats one combined megakernel even on first-token latency because the combined entry spills registers to local memory, which makes each prefill layer about $15\%$ slower in the planner's step benchmarks ($316$ vs $276$ $\mu s$ at B1).
\input{figures/original/figure-6-float}

Additive scoring sums event service and dispatch costs. The no-overlap ablation serializes otherwise concurrent work in a fixed topological order while preserving component costs and CTA budgets. Regret is selected measured latency divided by the measured minimum, minus one. Tied predictions do not count as unique-winner matches and receive their worst-member regret.

\section{How Far Should Kernel Fusion Extend?}
Fusion can stop anywhere between two extremes: conventional graph execution and monolith megakernel.

Making a region persistent adds a cost F that is fixed per step. Against this, the region saves S(b) relative to the conventional kernels it replaces at batch size b. Launch and host dispatch that a CUDA graph has not already removed are independent of b. Intermediate traffic that no longer reaches memory grows with b. The service difference between one compiled schedule and individually tuned kernels can grow, shrink or change sign as tiles fill and the number of waves changes. A region pays when $S(b)>F$. As $F$ is fixed while $S(b)$ can rise, fall or stay flat, the best extent of fusion moves with batch size and in different directions for different models.

This results in two observed categories of models, and two are predicted. For the two unobserved categories:
\begin{itemize}
\item Megakernel-advancing: We expect it where conventional cost is dominated by work-proportional traffic or by serialized compute, memory and communication that a persistent schedule can overlap.
\item Megakernel-averse: We expect it for regions already compute-bound such as large-model prefill component.
\end{itemize}

\noindent A region that gains little from a megakernel in isolation may nevertheless benefit when colocated, as the Nemotron experiment illustrates.

The answer to ``how far'' therefore ranges from multiple megakernels to conventional execution with thousands of kernels per phase. It depends on the model architecture, the batch size, and the objective. The boundaries worth preserving are those whose crossing costs less than the specialization and publication control they protect.

\input{tables/original-categories}

A model is megakernel-stable when its stable kernels:
\begin{itemize}
\item Account for most of the time on the path to the objective. For first-token latency that path is only prefill, which is why Llama-1B is stable for first-token latency but receding for decode.
\item Form contiguous regions whose savings exceed their fixed costs at every batch size. Stable kernels interleaved one by one with others can't pay for the cuts around them.
\item Do not lose their gains to co-located conventional parts, as in Observation 2.
\end{itemize}

\section{Related Work}
Persistent kernels are a long-standing GPU programming style~\cite{gupta2012persistent}. Hand-built megakernels for Llama-1B decoding and tensor-parallel Llama-70B~\cite{spector2025wholegpu,spector2025nobubbles}, compilers such as MPK~\cite{cheng2026mpk} and Event Tensor~\cite{jin2026eventtensor} and systems such as Ada-MK~\cite{dong2026adamk}, TileRT~\cite{tilert2026} and FlashMoE~\cite{aimuyo2025flashmoe} build fast megakernels for the regions they are given; we decide which regions those should be.

Rammer~\cite{ma2020rammer} selects persistent-program boundaries, while VersaPipe~\cite{zheng2017versapipe} tunes pipeline stage groups, execution models and resource assignments. Our focus is first-publication-aware composition using conventional and operator-level evidence. GPU performance forecasting models~\cite{lee2025forecasting} and interference-aware sharing systems~\cite{strati2024orion} provide complementary approaches; we combine service estimates with persistent resource lifetimes and streaming publication dependencies.

\section{Conclusion}
Fusing a streaming model into megakernels is a question of where to cut, not whether to fuse. Each cut decides which phases share a launch configuration, how long thread blocks stay held, how the phases share the GPU and which fixed costs are paid and the best answer ranges from zero to four megakernels across models and batch sizes. We showed that this answer can be predicted without timing candidate composition end to end, from the conventional baseline and standalone benchmarks of the operators involved. It selected the measured best design in all 13 cases across three models and three frameworks. The design it selects for NH-TTS achieves a 4.05\texttimes{} speedup in engine-level mean batch-p95 TTFA at batch 32 and the four rules derived from its model explain when fusion helps and when it does not.

These results open two directions. First, are megakernel-stable architectures universally superior to other models with comparable accuracy in terms of compute performance? Second, can a software-level scheduling prevent the interference between a co-located conventional graph and megakernel in GPU?

\bibliographystyle{bibliography/IEEEtran}
\bibliography{bibliography/IEEEabrv,bibliography/references}
\end{document}

%% file: figures/original/figure-1-float.tex
\begingroup
\setlength{\intextsep}{5pt}
\begin{figure}[H]
\centering
\includegraphics[width=\linewidth]{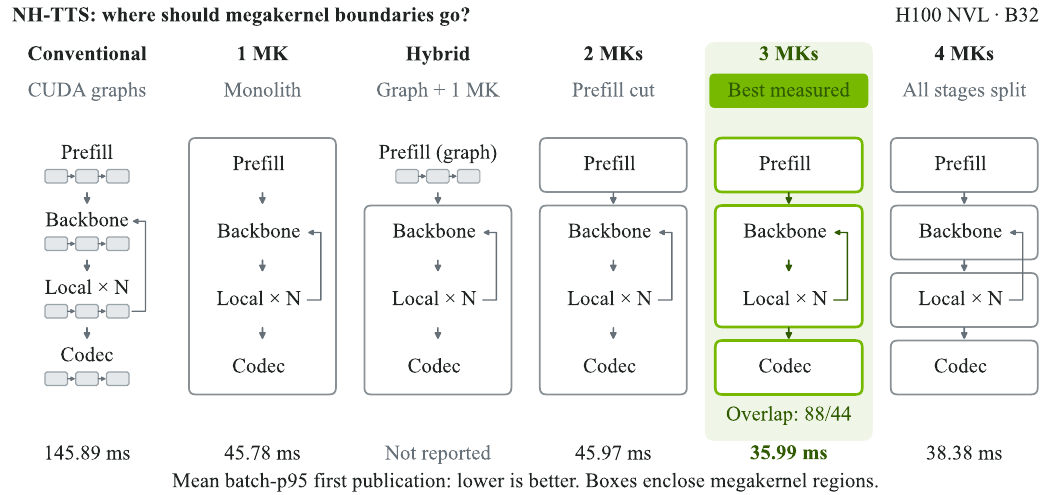}
\caption{Six representative NH-TTS layouts from the 81-candidate design space. The three-megakernel design with the lowest measured first-publication latency at batch 32 on H100 NVL is highlighted.}
\label{fig:nh_tts_layout}

\par\vspace{31.475pt}
\includegraphics[width=\linewidth]{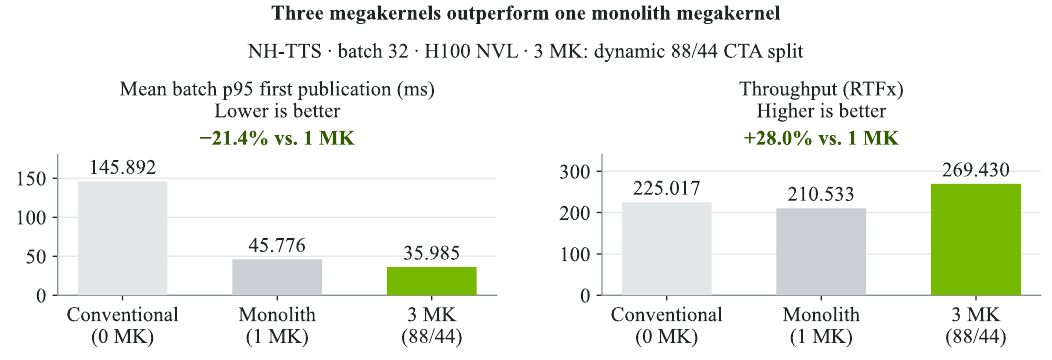}
\caption{TTFA and RTFx for NH-TTS at batch size 32 with the selected three-megakernel design}
\label{fig:mk_perf}
\end{figure}
\endgroup

%% file: figures/original/figure-4-float.tex
\begin{figure}[!t]
\centering
\includegraphics[width=.8\linewidth]{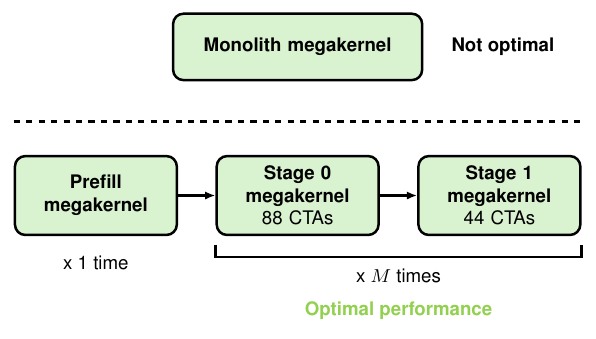}
\caption{Best megakernel composition for NH-TTS at batch size 32}
\label{fig:composition}
\end{figure}

%% file: tables/original-features.tex
\begin{table}[!htbp]
\caption{Feature modeling and algorithmic changes.}
\label{tab:features}
\centering
\footnotesize
\setlength{\tabcolsep}{4pt}
\renewcommand{\arraystretch}{1.1}
\begin{tabular}{|>{\centering\arraybackslash}m{1.9cm}|>{\centering\arraybackslash}m{.8cm}|m{4.7cm}|}
\hline
\rowcolor[HTML]{B2BEB5}\textbf{Feature} & \textbf{Modeled in} & \textbf{Algorithmic change} \\
\hline
Cross-kernel readiness & Tile graph & Edges: drop the barrier edges between consecutive kernels; keep the tile-to-tile tensor edges \\
\hline
Weight prefetch & Tile graph & Scheduling rule: one staging slot per worker. A staged tile's weight load is queued while its predecessors still run; its compute still waits for them \\
\hline
On-chip forwarding & Tile graph & Node weights: remove the forwarded tensor's store and reload from the memory queue \\
\hline
Producer--consumer overlap & Event graph & Edges, capacity: the first chunk depends on step $r$ instead of the preceding task, and runs in its own CTA allocation \\
\hline
Fewer launches, internal synchronization & Event graph & Nodes: a dispatch node before each launch, weighted by the per-launch cost; synchronization adds to tile weights \\
\hline
Smaller CTA budget & Service table & Tile scheduling uses $W$ workers. For an opaque operator with a grid of $g$ CTAs, compute time is approximated by scaling with $\lceil g/W\rceil/\lceil g/132\rceil$, assuming equal-cost CTA waves. \\
\hline
CTAs held while waiting & Event graph & Edges, capacity: a launch's release node depends on all its tasks; a launch is admitted only when its CTAs are free \\
\hline
Register pressure inside the fused entry & Service table & Node weights: each task's standalone time at the region's register limit, which is the register file divided by the resident threads per SM, at most 255 \\
\hline
Shared thread count \& tile shape & Service table & Node weights: pick one geometry per region, the one minimizing its summed task weights, and weight every task at it \\
\hline
Bandwidth contention & Event graph & Rates: each running task progresses at min(1, capacity $\div$ total demand) \\
\hline
\end{tabular}
\end{table}

%% file: algorithms/original-simulate.tex
\begin{algorithm}[H]
\caption{\textsc{Simulate}$(G,T,L,C)$}
\label{alg:simulate}
\small
\setlength{\parindent}{0pt}
\setlength{\parskip}{0pt}
\textbf{Require:}\par
\hspace*{1.7em}$G=(V,E)$: event graph\par
\hspace*{1.7em}$T[v]$: service time of node $v$ when it runs alone.\par
\hspace*{1.7em}$\mathrm{launch}[v]$: launch contract of node $v$, where for a launch $L$:\par
\hspace*{2.7em}$\mathrm{cta}[L]$: CTAs that L holds;\par
\hspace*{2.7em}$\mathrm{max\_tasks}[L]$: tasks of L that can run at same time;\par
\hangindent=1.7em\hangafter=0\hspace*{1em}$\mathrm{release}[L]$: control node that follows all tasks of L; its completion frees cta[L];\par
\hspace*{1.7em}$C$: shared memory bandwidth, where\par
\hspace*{2.7em}$B_{\mathrm{mem}}$: memory bandwidth of the device;\par
\hspace*{2.7em}$N_{\mathrm{CTA}}$: number of CTAs in the system;\par
\hspace*{2.7em}$Q[v]$: memory demand of node $v$.\par
\textbf{Ensure:} $\mathrm{finish}[v]$: completion time of each node $v$, or \textsc{Deadlock}.\par
\hspace*{1.7em}\textbf{State:}\par
\hspace*{2.7em}$\mathit{time}$: current simulated time; $T_{\mathrm{rem}}[v]$: time node $v$ still needs;\par
\hspace*{2.7em}$\mathrm{rate}[v]$: relative speed of $v$ (1 when standalone);\par
\hangindent=1.7em\hangafter=0\hspace*{1em}$\mathrm{READY}$: queue of unstarted nodes whose predecessors are all in $\mathrm{DONE}$;\par
\hspace*{2.7em}$\mathrm{RUNNING}$: nodes started but not finished;\par
\hspace*{2.7em}$\mathrm{DONE}$: finished nodes;\par
\hspace*{2.7em}$\mathrm{ADMITTED}$: launches currently holding CTAs;\par
\hspace*{2.7em}$\mathit{free\_cta}$: CTAs not held by an admitted launch.\par
\vspace{9pt}
\begin{algorithmic}[1]
\State $\mathit{time}\gets0$; \quad $T_{\mathrm{rem}}[v]\gets T[v]$ for every $v\in V$
\State $\mathrm{RUNNING}\gets\varnothing$; \quad $\mathrm{DONE}\gets\varnothing$; \quad $\mathrm{ADMITTED}\gets\varnothing$; $\mathit{free\_cta}\gets N_{\mathrm{CTA}}$
\State $\mathrm{READY}\gets\{\,v\in V:v\text{ has no predecessors}\,\}$
\While{$\mathrm{DONE}\ne V$}
  \ForAll{$v\in\mathrm{READY}$}
    \State $L\gets\mathrm{launch}[v]$
    \If{$L\ne\textsc{None}$ and $L\notin\mathrm{ADMITTED}$ and $\mathit{free\_cta}\ge\mathrm{cta}[L]$}
      \State $\mathrm{ADMITTED}\gets\mathrm{ADMITTED}\cup\{L\}$
      \State $\mathit{free\_cta}\gets\mathit{free\_cta}-\mathrm{cta}[L]$
    \EndIf
    \If{$L=\textsc{None}$ or $(L\in\mathrm{ADMITTED}$ and $|\{u\in\mathrm{RUNNING}:\mathrm{launch}[u]=L\}|<\mathrm{max\_tasks}[L])$}
      \State move $v$ from $\mathrm{READY}$ to $\mathrm{RUNNING}$
    \EndIf
  \EndFor
  \If{$\mathrm{RUNNING}=\varnothing$}
    \State \Return \textsc{Deadlock}
  \EndIf
  \State $Q_{\mathrm{tot}}\gets\sum_{v\in\mathrm{RUNNING}}Q[v]$
  \ForAll{$v\in\mathrm{RUNNING}$}
    \If{$Q[v]=0$}
      \State $\mathrm{rate}[v]\gets1$
    \Else
      \State $\mathrm{rate}[v]\gets\min(1,B_{\mathrm{mem}}/Q_{\mathrm{tot}})$
    \EndIf
  \EndFor
  \State $\Delta\gets\min_{v\in\mathrm{RUNNING}}T_{\mathrm{rem}}[v]/\mathrm{rate}[v]$
  \State $\mathit{time}\gets\mathit{time}+\Delta$
  \ForAll{$v\in\mathrm{RUNNING}$}
    \State $T_{\mathrm{rem}}[v]\gets T_{\mathrm{rem}}[v]-\mathrm{rate}[v]\cdot\Delta$
  \EndFor
  \ForAll{$v\in\mathrm{RUNNING}$ with $T_{\mathrm{rem}}[v]=0$}
    \State $\mathrm{finish}[v]\gets\mathit{time}$
    \State move $v$ from $\mathrm{RUNNING}$ to $\mathrm{DONE}$
    \If{$v=\mathrm{release}[L]$ for some launch $L$}
      \State $\mathrm{ADMITTED}\gets\mathrm{ADMITTED}\setminus\{L\}$
      \State $\mathit{free\_cta}\gets\mathit{free\_cta}+\mathrm{cta}[L]$
    \EndIf
    \ForAll{successors $w$ of $v$ in $G$ whose predecessors are all in $\mathrm{DONE}$}
      \State append $w$ to $\mathrm{READY}$
    \EndFor
  \EndFor
\EndWhile
\State \Return $\mathrm{finish}$
\end{algorithmic}
\end{algorithm}

%% file: algorithms/original-select-composition.tex
\begin{algorithm}[H]
\caption{\textsc{SelectComposition}$(R,b,E)$}
\label{alg:select}
\small
\setlength{\parindent}{0pt}
\setlength{\parskip}{0pt}
\textbf{Require:}\par
\hangindent=1.7em\hangafter=0$R$: region graph; one node per group; edges for data, state, and publication dependencies.\par
\hspace*{1.7em}$b$: batch size.\par
\hangindent=1.7em\hangafter=0$E$: profiles (kernel durations, HBM traffic, launch configuration), service records (measured or tile-simulated durations), and implementation contracts (backend, launch configuration, CTA budget, tile geometry).\par
\hangindent=1.7em\textbf{Ensure:} $\mathrm{BEST}$: set of feasible candidates with minimum first-publication score.\par
\vspace{10pt}
\hspace*{1.7em}\textbf{State:}\par
\hangindent=1.7em\hangafter=0\hspace*{1em}$\mathrm{CANDIDATES}$: supported candidates $\pi=(\mathrm{cover},\mathrm{policy})$ for $R$ at batch size $b$;\par
\hangindent=1.7em\hangafter=0\hspace*{2em}\textit{Cover}: split of the ordered groups into contiguous regions, each run by one backend;\par
\hangindent=1.7em\hangafter=0\hspace*{2em}\textit{Policy}: how regions share the GPU (serial, fixed split of CTAs, or dynamic split);\par
\hangindent=1.7em\hangafter=0\hspace*{1em}$\kappa[\pi]$: implementation context of $\pi$ (thread counts, tile geometry, call and storage policies);\par
\hangindent=1.7em\hangafter=0\hspace*{1em}$S[\pi]$: service cost of every group of $\pi$ for its backend, CTA budget, and context $\kappa[\pi]$;\par
\hangindent=1.7em\hangafter=0\hspace*{1em}$T_{\mathrm{pub}}[\pi]$: first-publication time of $\pi$ (when the first output can be returned).\par
\vspace{8.2pt}
\begin{algorithmic}[1]
\State $\mathrm{CANDIDATES}\gets$ every cover of $R$ paired with each applicable execution policy and supported contracts in $E$ at batch size $b$
\ForAll{$\pi\in\mathrm{CANDIDATES}$}
  \State $\kappa[\pi]\gets$ implementation context of $\pi$, from the contracts in $E$
  \State $S[\pi]\gets$ service costs from the records in $E$ whose context matches $\kappa[\pi]$
  \If{some group of $\pi$ has no matching record}
    \State $T_{\mathrm{pub}}[\pi]\gets\infty$; \textbf{continue}
  \EndIf
  \State $(G,t,L,C)\gets$ event graph of $\pi$: its tasks, launches, dispatch and release nodes, with times $t$ from $S[\pi]$, launch contracts $L$ from its regions and policy, bandwidth demands $C$, and a publication node that waits for the first consumer chunk and producer step $h$
  \State $\mathrm{finish}\gets\textsc{Simulate}(G,t,L,C)$
  \If{$\mathrm{finish}=\textsc{Deadlock}$}
    \State $T_{\mathrm{pub}}[\pi]\gets\infty$
  \Else
    \State $T_{\mathrm{pub}}[\pi]\gets\mathrm{finish}[\mathrm{publication}]$
  \EndIf
\EndFor
\State $T_{\mathrm{best}}\gets\min_{\pi\in\mathrm{CANDIDATES}}T_{\mathrm{pub}}[\pi]$
\State \Return $\mathrm{BEST}\gets\{\,\pi\in\mathrm{CANDIDATES}:T_{\mathrm{pub}}[\pi]=T_{\mathrm{best}}\text{ and }T_{\mathrm{best}}<\infty\,\}$
\end{algorithmic}
\end{algorithm}

%% file: tables/original-codec.tex
\begin{table}[!htbp]
\caption{Megakernel vs TensorRT Latencies}
\label{tab:latency_comparison}
\vspace{6pt}
\centering
\footnotesize
\setlength{\tabcolsep}{4pt}
\renewcommand{\arraystretch}{1}
\begin{tabular}{|c|c|c|c|}
\hline
\multicolumn{2}{|c|}{\cellcolor[gray]{.75}} & \multicolumn{2}{c|}{\cellcolor[gray]{.75}\textbf{Stage 1}$^a$} \\
\cline{3-4}
\multicolumn{2}{|c|}{\cellcolor[gray]{.75}} & \textbf{Megakernel} & \textbf{TensorRT} \\
\hline
\cellcolor[gray]{.75} & \textbf{Megakernel} & $0.991ms$ & $1.290ms$ \\
\cline{2-4}
\cellcolor[gray]{.75}\multirow{-2}{*}{\textbf{Stage 0}$^a$} & \textbf{TensorRT} & $1.216ms$ & $9.760ms$ \\
\hline
\multicolumn{4}{l}{$^{\mathrm{a}}$Refer Figure~\ref{fig:composition}} \\
\end{tabular}
\end{table}

%% file: tables/original-models.tex
\begin{table}[!htbp]
\caption{The three models in our evaluation.}
\label{tab:evaluation_models}
\centering
\footnotesize
\setlength{\tabcolsep}{4pt}
\renewcommand{\arraystretch}{1}
\begin{tabular}{|>{\raggedright\arraybackslash}p{1cm}|>{\raggedright\arraybackslash}p{2.8cm}|p{2cm}|>{\raggedright\arraybackslash}p{1.2cm}|}
\hline
\rowcolor[gray]{.75}\textbf{Model} & \textbf{Conventional baseline} & \mbox{\textbf{First publication}} & \mbox{\textbf{Batch sizes}} \\
\hline
NH-TTS & TensorRT-LLM with CUDA graphs, plus TensorRT engines & Time to First Audio (TTFA) at engine publication & 1, 32, 64 \\
\hline
Qwen3-TTS & PyTorch with vLLM FlashAttention-3, torch.compile and CUDA graphs & Time to First Audio (TTFA) on the host & 1, 16, 32, 48, 64 \\
\hline
$^{\mathrm{a}}$Llama 3.2 1B & PyTorch SDPA (FlashAttention) in one CUDA graph & Time to First Token (TTFT) & 1, 2, 16, 32, 48 \\
\hline
\multicolumn{4}{l}{$^{\mathrm{a}}$496-token prompt} \\
\end{tabular}
\end{table}

%% file: figures/original/figure-6-float.tex
\begin{figure}[!htbp]
\centering
\includegraphics[width=\linewidth]{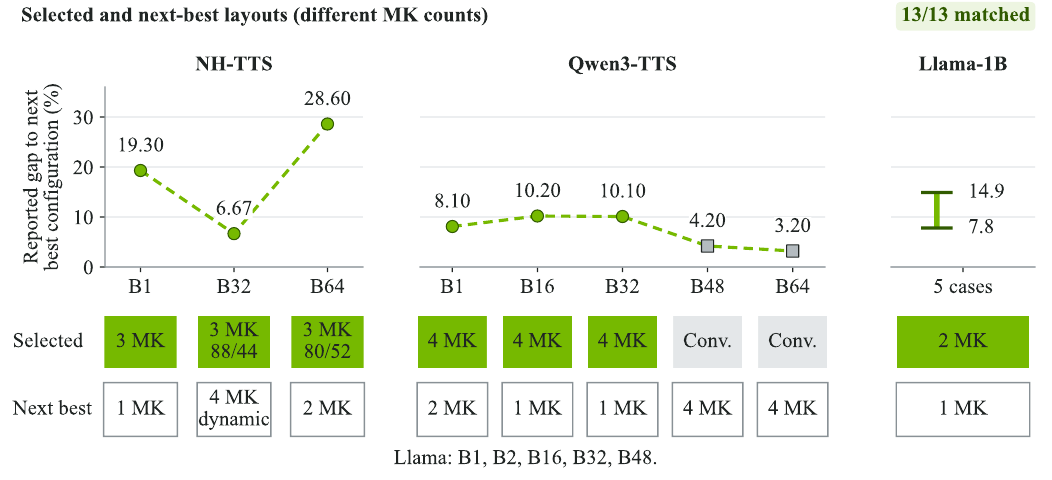}
\caption{Predicted selections within the implemented validation pools. Predicted and measured margins compare the winner with the nearest alternative in the same pool.}
\label{fig:selection}
\end{figure}

%% file: tables/original-categories.tex
\begin{table}[!htbp]
\caption{Four model categories by where megakernels pay}
\label{tab:model_categories}
\vspace{6pt}
\centering
\footnotesize
\setlength{\tabcolsep}{6pt}
\renewcommand{\arraystretch}{1}
\begin{tabular}{|m{.5cm}|m{.8cm}|m{2.8cm}|m{2.8cm}|}
\hline
\multicolumn{2}{|l|}{\cellcolor[gray]{.75}} & \multicolumn{2}{c|}{\cellcolor[gray]{.75}\textbf{Large}} \\
\cline{3-4}
\multicolumn{2}{|l|}{\cellcolor[gray]{.75}\multirow{-2}{*}{\textbf{\ Batch}}} & \textbf{\textit{Pays}} & \textbf{\textit{Doesn't}} \\
\hline
\cellcolor[gray]{.75} & \textbf{\textit{Pays}} & Megakernel-stable (NH-TTS; Llama-1B for first-token latency) & Megakernel-receding (Qwen3-TTS; Llama-1B for inter-token latency) \\
\cline{2-4}
\cellcolor[gray]{.75}\multirow{-3}{*}{\textbf{Small}} & \textbf{\textit{Doesn't}} & Megakernel-advancing$^{\mathrm{a}}$ & Megakernel-averse$^{\mathrm{a}}$ \\
\hline
\multicolumn{4}{l}{$^{\mathrm{a}}$Predicted, not observed.} \\
\end{tabular}
\vspace{6pt}
\end{table}